\documentstyle[prb,aps,multicol,graphicx]{revtex}  

\tighten       

\begin{document}

\draft                          


\title{Raman spectroscopy of InN films grown on Si}

\author{F.~Agull\'o-Rueda%
\thanks{Electronic mail: far@icmm.csic.es}
}

\address{Instituto~de~Ciencia~de~Materiales~de~Madrid (CSIC), 
Cantoblanco, E-28049~Madrid,~Spain}

\author{E.~E.~Mendez
       }

\address{Department of Physics and Astronomy, 
         State University of 
         New York at Stony Brook,
         Stony Brook, NY 11794-3800, USA
         }

\author{N.~Bojarczuk and S.~Guha
       }

\address{IBM Research Division, 
         IBM Thomas J. Watson Research Center, 
         P.\-O. Box 218,
         Yorktown Heights, NY 10598, USA}

\date{Received * * *
}

\maketitle


\begin{abstract}   
   We have used Raman spectroscopy    
   to study indium nitride
   thin films grown by molecular 
   beam epitaxy on (111) silicon substrates
   at temperatures between
   450$^\circ$ and 550$^\circ$C.
   The Raman spectra show well
   defined peaks at 443, 475, 491, and 591~cm$^{-1}$,
   which correspond to the
   $A_1$(TO), $E_1$(TO), $E_2^{\rm high}$, and $A_1$(LO)
   phonons of the wurtzite structure, respectively.
   In backscattering normal to the surface
   the $A_1$(TO) and $E_1$(TO) peaks are very weak,
   indicating that the films grow 
   along 
   the hexagonal $c$ axis.
   The dependence of the peak width on growth 
   temperature reveals that the optimum 
   temperature is 500$^\circ$C, 
   for which the fullwidth of the 
   $E_2^{\rm high}$ peak has the minimum value of 
   7~cm$^{-1}$.     
   This small value, comparable to previous
   results for InN films grown on sapphire,
   is evidence of the good crystallinity of the films.
\end{abstract}





\begin{multicols}{2}



The III-V direct-gap semiconductor 
InN has been largely
ignored because its low dissociation
temperature makes it very 
difficult to grow. 
However, in the last few years there has been an
increasing interest on the material 
due, to large extent, to
the successful application
of nitride compounds in ultraviolet-blue
light-emitting diodes 
and lasers. 
In particular,
InN has promising transport 
and optical properties.
Its large drift velocity
at room temperature
could render it better than GaAs and GaN
for field effect transistors.\cite{Oleary98}
Carrier capture by InN quntum dots
has been adduced
for the efficient emission
of blue-violet commercial InGaN diode lasers.%
\cite{Odonnell99}
InN/Si tandem solar cells
have been proposed for increased efficiency.%
\cite{Yamamoto94}
Finally, the important quaternary 
alloy AlGaInN covers most of the visible spectrum,
reaching the orange-red end for InN.

Due to the lack of suitable lattice-matched
substrates, InN thin films have been grown 
mostly on sapphire,
which is widely available. 
Although silicon would be preferable
for device applications and integration
with microelectronic integrated circuits,
films grown directly on Si substrates
are poorly oriented.\cite{Yamamoto94b}
The reason is that In adatoms have a
long migration length that causes the
formation of InN islands 
during the initial stages.
The exposed Si surface reacts with the
nitrogen beam to produce
amorphous SiN, hindering the
growth of high quality InN.%
\cite{Yamamoto94b,Bello92}
An AlN buffer layer has been shown
to improve the quality of InN films
grown on sapphire\cite{Kistenmacher91}
and also of GaN films on Si.\cite{Watanabe93}
Since Al has a short migration length,
a thin, uniform AlN layer can be 
grown, avoiding the reaction of the substrate
with the nitrogen beam.

As other nitrides, InN can
crystallize with the wurtzite hexagonal 
or the zincblende cubic structures.
Raman spectroscopy has 
been extensively used
to determine the structure and
the crystallinity of GaN. 
For InN, previous work has been limited to
wurtzite films grown on (0001) 
sapphire\cite{Kwon96,Lee98,Inushima99}
and zincblende films grown on (001) GaAs.\cite{Tabata99}


In this paper we report the
growth and characterization by Raman
scattering of oriented, 
crystalline InN films. 
InN thin films were grown by
molecular beam epitaxy on (111) 
Si substrates using a RF 
plasma nitrogen source.
Three different samples were 
grown at substrate temperatures 
$T_{\rm g}$ of
450$^\circ$, 500$^\circ$ and 550$^\circ$C, respectively.
A 10-nm-thick AlN buffer layer was
deposited between the substrate and the
InN layer.  
Under the optical microscope the films
presented a domain-like morphology.
The domain size increased with
growth temperature. 
The film grown 
at 550$^\circ$C
showed a poor adherence to the substrate.

Room temperature
Raman spectra were taken 
with a Renishaw Ramascope
spectrometer,
equipped with an Ar$^+$ ion laser 
as a light source
operating at a
wavelength of 514.5~nm
and focused on the sample through an optical
microscope.
The power density on the surface 
was of the order of
100~kw/cm$^2$.
Light scattered by the sample was
collected with the same microscope
and analyzed with a single-grating
spectrograph and a CCD detector.
Reflected and elastically scattered light was
blocked with two holographic filters,
which also removed most of the Raman spectrum
below 100~cm$^{-1}$.



\begin{figure}
      \includegraphics[width=8.5cm]{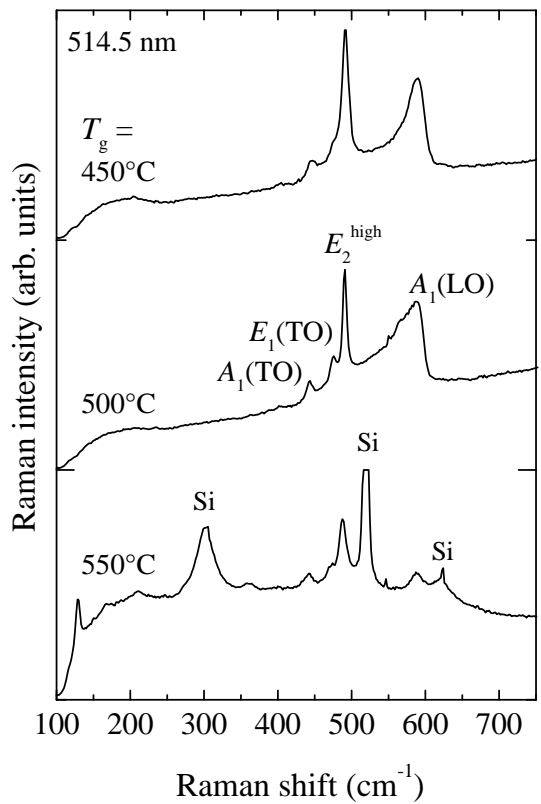}
   \caption{
   Raman spectra for the InN thin films 
   grown at three substrate temperatures $T_{\rm g}$.    
   The spectra have been shifted vertically 
   for clarity.
   Ticks on the vertical axis mark the zero
   level for each case.
   Peaks originating in the substrate 
   are labelled as Si.
   \label{fig:Raman}
   }     
\end{figure}

Figure~\ref{fig:Raman} shows the
Raman spectra for the InN films
grown at different temperatures.
All samples exhibit four peaks
characteristic of bulk InN.
In addition,
the 550$^\circ$C sample shows some
peaks originating from the silicon
substrate (labelled Si in the figure),
due to its deficient coverage.
The other samples show no signal
from the substrate, and none of 
the three spectra reveal
any band coming
from the AlN buffer 
layer.\cite{Mcneil93}


The zincblende structure
(spatial group $T_d^2 - F\bar{4}3m$)
has only two Raman-active 
phonons $F_2$(TO) and $F_2$(LO).
The wurtzite structure 
(spatial group $C_{6v}^4 - P6_3mc$),
which is the most stable,
has six Raman-active phonons,
$A_1$(TO), $A_1$(LO), $E_1$(TO), 
$E_1$(LO), and $2E_2$.
Therefore the number of peaks
observed in the spectra 
indicates that the wurtzite
phase must be present.


Table~\ref{tab:Raman} lists the
positions of the peaks and their
symmetry assignment.
The strongest peaks correspond
to the $E_2^{\rm high}$ phonon 
at around 491~cm$^{-1}$ and
to the $A_1$(LO) phonon at
around 591~cm$^{-1}$.
The frequency of the $E_2^{\rm high}$ phonon
is very close to the value of 488~cm$^{-1}$ 
reported for InN on
sapphire.\cite{Inushima99}

The peaks at 443 and 475~cm$^{-1}$
have been identified
as the $A_1$(TO) and the $E_1$(TO) phonons,
respectively,
of the hexagonal phase.
This assignment has been done by
comparison with the intensities of 
good quality GaN Raman spectra 
and by following the trend of phonon
frequencies
in AlN and GaN.\cite{Mcneil93,Ponce96} 
The frequency of the $A_1$(TO) phonon
agrees well with the value of 450~cm$^{-1}$
calculated by Kim {\em et al},\cite{Kim96}
but is much lower than the value reported
by Inushima {\em et al},\cite{Inushima99}
who assigned it to a shoulder in the
Raman spectrum at 480~cm$^{-1}$.
On the other hand, 
the frequency of the $E_1$(TO) phonon
coincides with the value of 
Inushima {\em et al},\cite{Inushima99}
but disagrees with the value of
580~cm$^{-1}$ estimated 
by Kim {\em et al}.\cite{Kim96}
Both the $E_1$(TO) and the $A_1$(TO)
are forbidden for backscattering
along the hexagonal $c$ axis.
The fact that these peaks are very weak
indicates that the films grow with 
a preferential orientation of this
axis normal to the substrate plane.
   
The $A_1$(LO) peak at 591~cm$^{-1}$
shows a low energy tail whose
intensity depends on the sample and even
on the measuring point.
It could arise from the 
appearance of the forbidden
mode $E_1$(LO), which has been
reported\cite{Inushima99}
to be at 570~cm$^{-1}$.
Alternatively, the low-energy tail can be attributed 
to LO-phonon-plasmon coupling
due to residual free carriers, whose concentration
increases with increasing growth temperature.
This interpretation would explain the observed increase of 
the continuous background
with temperature, an effect that has
been associated in GaN with
an increasing dopant density.\cite{Harima98}


The full-width at
half maximum (FWHM)
of the $E_2^{\rm high}$
Raman peak, which varies from 
13 to 7 ~cm$^{-1}$ 
(see Table~\ref{tab:Raman}),
depends on the
crystallinity of the films.
Usually, the peak is broadened by 
reduced phonon coherence caused
by lattice disorder,
the formation of nanocrystals or
the presence of defects and
impurities.
The observed linewidth indicates that the
best crystallinity is obtained 
for a growth temperature of 500$^\circ$C.
The FWHM value of 7~cm$^{-1}$ at this temperature is
comparable
to the value of 5~cm$^{-1}$ observed
for the best films grown on sapphire.\cite{Lee98,Pan99}
Although the latter has been preferred
among the lattice-mismatched
substrates, our results show that
buffered silicon can also produce 
InN films with good crystallinity. 
This offers the advantage of a better
compatibility with microelectronic
integrated circuits and other silicon-based
devices.


In summary, by depositing a
thin AlN buffer layer we have been able to grow
InN films with good crystallinity
on Si substrates by
molecular beam epitaxy.
An analysis of their
Raman spectra
show that the films have a wurtzite
structure with the hexagonal
$c$ axis perpendicular to the substrate plane.
The best crystallinity of the samples is achieved 
for a growth temperature of (or close to) 500$^\circ$C.


   We acknowledge financial support from the Spanish 
   CICyT (Projects MAT96-0395-CP and MAT97-0725)
   and the US Army Research Office.
   We thank C. Pecharrom\'an for
   helpful discussions.

\begin{table}

   \caption{
   Summary of Raman 
   phonon frequencies in cm$^{-1}$
   for the InN thin films grown on Si at different 
   substrate temperatures $T_{\rm g}$.
   The full widths of the peaks at half maximum are
   given in parenthesis in cm$^{-1}$.
   The symmetry species are for the wurtzite
   structure.
   \label{tab:Raman}
   }

   \begin{footnotesize}

   \begin{tabular}{l l l l}

      $T_{\rm g}$                                                 
      ($^\circ$C) &  450     & 500     & 550        \\ \tableline                                                      
          
      $A_1$(TO)   &  445(17) & 443(10) & 441(14)    \\            
      $E_1$(TO)   &  475(17) & 475(8)  & 471(14)    \\                                                                 
      $E_2^{\rm high}
                  $       &  491(10) & 491(7)  & 489(13)    \\    
      $A_1$(LO)   &  590     & 591     & 588        

      \end{tabular}

      \end{footnotesize}         

\end{table}

%
%

\end{multicols}

\end{document}